\documentclass[reprint,superscriptaddress,aps,pra]{revtex4-2}
\usepackage{graphicx}
\usepackage[dvipsnames]{xcolor}
\usepackage[colorlinks=true,linkcolor=Blue,citecolor=Blue,urlcolor=Blue,linktocpage=true]{hyperref}
\usepackage{amsmath,amssymb,amscd,amsfonts,bm, bbm}
\usepackage{mathptm}

\begin{document}
	
\title{Force sensing in an optomechanical system with feedback-controlled in-loop light}

\author{F. Bemani}  
	\email{foroudbemani@gmail.com}
	\affiliation{Department of Optics, Palack\'{y} University, 17. listopadu 1192/12, 77146 Olomouc, Czechia}

\author{O. \v{C}ernot\'{i}k} 
	\email{ondrej.cernotik@upol.cz}
	\affiliation{Department of Optics, Palack\'{y} University, 17. listopadu 1192/12, 77146 Olomouc, Czechia}

\author{L. Ruppert} 
	\email{ruppert@optics.upol.cz}
	\affiliation{Department of Optics, Palack\'{y} University, 17. listopadu 1192/12, 77146 Olomouc, Czechia}

\author{D. Vitali}
	\email{david.vitali@unicam.it}
	\affiliation{Physics Division, School of Science and Technology, University of Camerino, I-62032 Camerino (MC), Italy}
	\affiliation{INFN, Sezione di Perugia, via A. Pascoli, Perugia, Italy}
	\affiliation{CNR-INO, L.go Enrico Fermi 6, I-50125 Firenze, Italy}

\author{R. Filip} 
	\email{filip@optics.upol.cz}
	\affiliation{Department of Optics, Palack\'{y} University, 17. listopadu 1192/12, 77146 Olomouc, Czechia}
		
\date{\today}
			
\begin{abstract} 
Quantum control techniques applied at macroscopic scales provide us with opportunities in fundamental physics and practical applications. Among them, measurement-based feedback allows efficient control of optomechanical systems and quantum-enhanced sensing. In this paper, we propose a near-resonant narrow-band force sensor with extremely low optically added noise in an optomechanical system subject to a feedback-controlled in-loop light. The membrane's intrinsic motion consisting of zero-point motion and thermal motion is affected by the added noise of measurement due to the backaction noise and imprecision noise. We show that, in the optimal low-noise regime, the system is analogous to an optomechanical system containing a near quantum-limited optical parametric amplifier coupled to an engineered reservoir interacting with the cavity. Therefore, the feedback loop enhances the mechanical response of the system to the input while keeping the optically added noise of measurement below the standard quantum limit. Moreover, the system based on feedback offers a much larger amplification bandwidth than the same system with no feedback.
\end{abstract}

\maketitle
		
\section{\label{sec:Sec1} Introduction} 
	Quantum control is a developing research area that explores techniques guiding the system dynamics towards desired targets by external, time-dependent manipulation. Advanced quantum control schemes are important when exploiting coherent quantum effects in the emerging field of quantum technologies. Feedback is a control technique where the outputs from the quantum system are used as inputs that control the system dynamics and which enables us to evolve the system towards some desired outcome. Feedback techniques show promise for system stabilization and reducing decoherence and noise \cite{Zhang,Serafini}. Various  optical demonstrations of feedback techniques have been shown over the past few years.  Prominent among these  are the coherent quantum feedback and measurement-based feedback control \cite{Zhang,Serafini}. The latter is a powerful platform for the state control of the mechanical oscillator and enhanced quantum sensing \cite{Zhang,Serafini}. 

 	Combining ultralow-dissipation optical cavities with advanced nanofabrication techniques for mechanical resonators has led to outstanding experimental progress in the past decade \cite{Aspelmeyer}. These efforts have resulted in the coherent interaction of light and mechanical motion of massive objects at the quantum level and paved the way for a better understanding of light-matter interaction in the rapidly developing area quantum optomechanics. Feedback techniques have also been used in optomechanical systems.  For instance, approaching the quantum ground state of a kilogram-scale system in a gravity-wave interferometer \cite{Abbott},   quantum-enhanced feedback cooling of a mechanical oscillator \cite{Schafermeier,Poggio}, and force and position sensing via measurement-based control of a mechanical oscillator \cite{Wilson,rossi2018measurement} have been investigated with a feedback loop that operates directly on the mechanical element. More recently, feedback-controlled in-loop light has been used to enhance the efficiency of optomechanical systems, e.g., enhancing sideband cooling \cite{Rossi2}, normal-mode splitting \cite{Rossi}, improving the performance of an optomechanical heat engine \cite{Abari}, and generating entanglement of two different mechanical resonators \cite{Jie}.
			
	Any measurement of an observable that does not commute with itself at different times has limits due to the Heisenberg uncertainty theorem \cite{Clerk2010}. The standard quantum limit (SQL) is determined by a trade-off between shot noise and radiation pressure backaction noise in force sensors based on quantum optomechanics. Various approaches have been proposed and experimentally realized to surpass the SQL: backaction-noise reduction to overcome the SQL \cite{Khalili,Caniard,Pontin,Lecocq},  coherent quantum noise cancellation based on quantum interference \cite{Bariani,Wimmer,Motazedifard_2016,Buchmann,Khalili2018,moller2017quantum} and exploitation of quantum correlations to deviate from the SQL \cite{Mason2019}. In force sensors based on the backaction-evasion technique, the SQL is surpassed by producing a large signal without suppressing the added noise while in the quantum noise cancellation technique, the backaction noise is completely cancelled without amplifying the signal. It has been also demonstrated that the parametric modulation of the spring coefficient of a mechanical oscillator in an optomechanical system could be used to perform single-quadrature detection of a force applied to the mechanical oscillator with simultaneously suppressed optically added measurement noise and amplified input signal  \cite{Levitan_2016}. 
	
	In the red detuning regime, the optomechanical interaction causes a net energy transfer from the mechanical mode into the cavity mode, hence, in effect, cools the mechanical motion \cite{Arcizet2006,Gigan2006,Schliesser2006,Wilson-Rae,Marquardt,Schliesser2008}. In contrast, in the blue detuning regime, the optomechanical interaction results in an energy transfer into the mechanical motion; consequently, it results in optomechanical amplification \cite{Massel2011,McRae2012,Kippenberg2005}. In this situation the mechanical noise is not cooled and induces a significant added noise. Amplification can also be realized in optomechanical systems subject to mechanical parametric driving \cite{Levitan_2016}. Both optical and mechanical responses of a linear amplifier for optomechanical systems to quantum and classical fluctuations have been developed in Ref.~\cite{Botter}.
	
	Inspired by the above-mentioned studies on feedback control and optomechanical force sensors, in this paper, we propose a high--precision near-resonance narrow-band force sensor that is feasible with state-of-the-art experimental setups \cite{Rossi,Rossi2}. We show a close analogy between an optomechanical system with in-loop feedback and an optomechanical amplifier with an engineered reservoir interacting with the cavity. Consequently, the feedback loop enhances the mechanical response of the system to the input force in the red detuning of the resolved sideband regime. The force sensor introduced here exhibits simultaneously \textit{quantum-limited signal amplification} (due to the feedback loop) and \textit{noise suppression} (due to the cooling by the red-sideband laser drive) in the resolved sideband regime. Moreover, the weak input signal is transduced with higher gain to the optical output while the signal-to-noise ratio and sensitivity of the system remain unchanged. The feedback loop also significantly enhances the detection bandwidth.
	
	The paper is organized as follows. In Sec. \ref{sec:Sec2}, we describe the system under consideration, i.e., a nanomechanical membrane placed inside an optical cavity subject to a feedback-controlled in-loop light, and then we present the quantum Langevin equations.  In Sec. \ref{sec:Sec3}, we demonstrate how to amplify the input signal by enhancing the mechanical response of the system by feedback-controlled in-loop light, at the same time suppressing the added noise of measurement by red detuned laser drive. Then, we study the sensitivity as well as the signal-to-noise ratio of the proposed force sensor. Finally, in Sec. \ref{sec:Sec4}, we present our concluding remarks.
	
\section{\label{sec:Sec2} Theoretical description of the system}
	\begin{figure}
		\includegraphics[width=8.6cm]{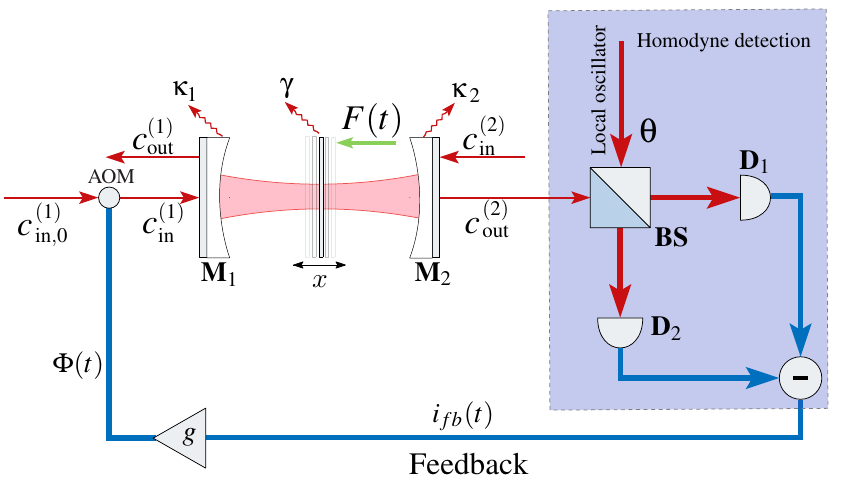}
		\caption{Schematic illustration of the proposed optomechanical force sensor: A nanomechanical membrane is positioned inside a high--finesse optical cavity with feedback-controlled in-loop light. The cavity is driven through the mirror ${\bf M}_1$ and the amplitude of the transmitted beam at mirror ${\bf M}_2$ is detected via homodyne detection which consists of a beam splitter ($ \bf BS $), two photodiodes ($ {\bf D}_1 $, $ {\bf D}_2 $) and a local oscillator with a phase $\theta$ relative to the signal. The resulting signal $ i_{\rm fb}(t)$ is electronically processed, and fed back, through an acousto-optic modulator (AOM) driver to  the amplitude of the input field at mirror ${\bf M}_1$, thus closing the feedback loop.}
		\label{Fig:Fig1}
	\end{figure}
	Our scheme for the feedback-enhanced force sensor is depicted in Fig.~\ref{Fig:Fig1}. A nanomechanical membrane is placed into a high--finesse double-sided  Fabry--Perot cavity with frequency $\omega_c$ which is driven through the mirror ${\bf M}_1$. Compared with a single-sided optomechanical cavity, this setup contains one additional parameter to optimize, allowing more efficient detection. In addition, the position of the membrane can be measured by monitoring the transmitted light using balanced homodyne detection. The resulting photocurrent, $ i_{\rm fb}(t)$, is fed back to the input at mirror ${\bf M}_1$ using an acousto-optic modulator (AOM) driver that modulates the amplitude of the input. The magnitude of an external force on the sensor is then estimated using standard linear response theory. We will show how to amplify the force by applying the feedback loop in the system.
	
	We assume a single mechanical mode coupled to another optical mode via radiation pressure \cite{Aspelmeyer}. We consider a strongly driven optical cavity field and weak optomechanical coupling. This allows us to linearize the quantum dynamics of fluctuations around the semiclassical amplitudes \cite{Aspelmeyer,Kippenberg}. The linearized optomechanical Hamiltonian describing the interaction is given by
	\begin{equation} \label{Hamiltonian1}  
		H =  \hbar \Delta c^\dag c + \hbar\omega_{\rm m} b^\dag b + \hbar \Lambda(c^\dag  + c)(b^\dag  + b)+{x_{{\rm{zpf}}}} F(t) (b^\dag  + b)\, .
	\end{equation}
	The first term describes the free energy of the cavity mode in a frame rotating at laser frequency where $c$ ($ c^\dag  $) is the annihilation (creation) operator of the cavity mode and $\Delta = \omega_{\rm c}-\omega_{\rm L}$ is the detuning between the cavity resonance $\omega_{\rm c}$ and the pump laser frequency $\omega_{\rm L}$. The second term describes the mechanical Hamiltonian where the operator $b$ ($ b^\dag  $) is the annihilation (creation) operator of the mechanical mode with natural frequency $\omega_{\rm m}$. The interaction between the center-of-mass motion of the membrane and the cavity mode is given by the third term in Eq.~(\ref{Hamiltonian1}) where $\Lambda$ is the interaction rate. Finally, the last term in the Hamiltonian accounts for the coupling of the membrane to an input classical force $F(t)$ in the $x$-direction, as depicted in Fig.~\ref{Fig:Fig1}. Moreover, $ x_{\rm zpf}=\sqrt{\hbar/{2 m \omega_m}}$ is the zero-point posotion uncertainty of the membrane with mass $ m$.

	Moving the operators to an interaction picture with respect to the Hamiltonian $H_0=\hbar \Delta c^\dag c + \hbar\omega_{\rm m} b^\dag b $, we obtain the linearized optomechanical Hamiltonian describing the interaction as
	\begin{equation}\label{Hamiltonian2}
		H_{\rm I} =  \hbar \Lambda(c^\dag b + b^\dag c) +{x_{{\rm{zpf}}}} F(t)\big(b^\dag e^{i \omega_{\rm{m}} t}  + be^{-i \omega_{\rm{m}} t}\big)+H_{\rm CR} \, .
	\end{equation}
	with $H_{\rm CR}=\hbar \Lambda [cb\exp(-2i \omega_{\rm{m}} t)+b^\dagger c^\dagger\exp(2i \omega_{\rm{m}} t)] $ where we have also assumed $\Delta \simeq \omega_m$.  In the \textit{resolved sideband regime} where the mechanical frequency is much larger than the damping rate of the cavity ($ \omega_{\rm{m}}\gg \kappa $), $ H_{\rm CR} $ can be approximately neglected. To account for the small correction to the dynamics caused by $ H_{\rm CR} $ in this regime,  we use the sideband truncation approach described in appendix \ref{Appendix:A} in our numerical calculations. 
\subsection{Applying the coherent feedback loop}
	To investigate the system dynamics, we now include fluctuation and dissipation processes affecting the optical and the mechanical modes by adding for each of them the corresponding damping and noise term and write the following quantum Langevin equations
	\begin{align}
		\dot b = & - \gamma  b + i\Lambda c +\sqrt {2\gamma } \, \, b_{ \rm in,f }(t) \,, \label{equations_of_motion1}\\
		\dot c =&  - \kappa \, c + i\Lambda b + \sqrt {2 \kappa _1} \,  c_{\rm in}^{(1)}(t) + \sqrt {2\kappa _2} \,  c_{\rm in}^{(2)}(t)\,,\label{equations_of_motion2}
	\end{align}
	where $ c_{\rm in}^{(1)}(t) $ and $ c_{\rm in}^{(2)}(t) $ are the noise operators associated with the mirrors $\bf{M}_1 $ and $ \bf{M}_2$, respectively. The nanomechanical membrane interacts with a thermal environment at a finite temperature with a damping rate of $\gamma$. The input mechanical operator $b_{\rm in,f}(t)$ consists of the external force to be detected, $ F(t) $, and a noise operator $  b_{\rm in }(t) $,
	\begin{equation}\label{Modified_mechanical_noise}
		b_{ \rm in,f }(t) = -\frac{i}{2\sqrt {\hbar m\gamma \omega _{\rm m}} }F(t)e^{i \omega_{\rm{m}} t} + b_{\rm in }(t)\,.
	\end{equation}
	Here, the mechanical noise $ b_{\rm in} $  affecting the nanomechanical membrane satisfies the Markovian correlation functions
	\begin{align}
		\langle b_{\rm in}(t) \, b_{\rm in}^\dag (t') \rangle&= (1+\bar n)\,  \delta (t-t')\,, \label{Mechanical_correlation1}\\
		\langle b_{\rm in}^\dag (t) \, b_{\rm in}(t') \rangle&= \bar n \, \delta (t-t')\,, \label{Mechanical_correlation2}
	\end{align}
	where $ \bar n=[\exp( \hbar \omega_{\rm m} /k_B T)-1]^{-1} $ is the mean number of thermal excitations of the mechanical bath at temperature $ T $.	Moreoevre, the nonzero noise correlators associated with the cavity input are given by \cite{Aspelmeyer} 
	\begin{align}
			\langle c_{\rm in,0}^{(1)} (t) \, c_{\rm in,0}^{{(1)} \dag}(t') \rangle&=  \delta (t-t')\,, \label{optical_correlation1}\\
	\langle c_{\rm in}^{(2)} (t) \, c_{\rm in}^{{(2)} \dag}(t') \rangle&=  \delta (t-t')\,, \label{optical_correlation2}
	\end{align}
By introducing the total cavity decay rate $ \kappa=\kappa_1+\kappa_2 $ in terms of the corresponding decay rates of the mirrors, we can formally write the total input noise operator for the cavity field as $ \sqrt {2\kappa} c_{\rm in}(t) = \sqrt {2 \kappa _1} \,  c_{\rm in}^{(1)}(t) + \sqrt {2\kappa _2} \,  c_{\rm in}^{(2)}(t) $. 
	
	We now modify the system dynamics by applying feedback-controlled in-loop light \cite{Abari,Rossi,Rossi2,Zippilli}. The operator $ c_{\rm out}^{(2)} $ describes the transmitted output field given by the input–output relation
	\begin{equation}\label{input_output}
		c_{\rm out}^{(2)}(t) =  \sqrt {2\kappa_2}\,  c(t) - c_{\rm in}^{(2)}(t).
	\end{equation}
	We then consider a phase-sensitive detection of an output quadrature at phase $ \theta $
	\begin{equation}\label{quadrature}
		X_{\rm out,fb}^\theta (t) = \frac{{{e^{ - i\theta }}c_{\rm out}^{(2)}(t) + {e^{i\theta }}c_{\rm out}^{(2)\dag }(t)}}{{\sqrt 2 }}\,.
	\end{equation}
	The resulting photocurrent at the output of the second mirror is expressed as \cite{Abari,Zippilli}
	\begin{equation}\label{photocurrent}
		i_{\rm fb}(t) = \sqrt {\eta}  \, X_{{\rm out,fb}}^{\theta}(t) + \sqrt {1 - \eta} \,  X_\nu (t),
	\end{equation}
	Here, we model the output light due to inefficient detection of the optical field where $ \eta $ is the detection efficiency and the operator $ {X_\nu } $ accounts for the additional noise due to inefficient detection and fulfills the relation $ \langle X_\nu ( t ) X_\nu(t') \rangle  =(1/2) \delta (t - t') $. 

	As shown in Fig.~\ref{Fig:Fig1}, the photocurrent is fed back using an AOM driver to modulate the amplitude of the input field at mirror $ {\bf M}_1 $.  The overall effect of the feedback loop can be modeled by the electronic noise $\Phi(t)$ modulating (displacing) the open-loop input noise operator $c_{\rm in,0}^{(1)}(t)$. Therefore, the input noise operator applied to the mirror $M_1$ is written as the superposition of the original input $  c_{{\rm{in,0}}}^{(1)}(t) $ and an additional term due to the feedback $ \Phi (t) $ according to 
	\begin{equation}\label{input_feedback_noise}
		c_{\rm in}^{(1)}(t) = c_{{\rm{in,0}}}^{(1)}(t) + \Phi (t).
	\end{equation}
	where $ \Phi (t) = g i_{\rm fb} (t - \tau)$, $ g $ is the feedback gain, and $\tau$ is the signal time delay in the loop. We consider a feedback loop with a non-zero delay caused by a latency of electronic processing. For high--quality electronic circuits, we can neglect all dispersion effects.  
Therefore, we write the delay-time dependence of the photocurrent in Eq.~(\ref{photocurrent}) as
	\begin{align}\label{photocurrent2}
		i_{\rm fb}(t - \tau ) =& \sqrt {\eta/2}\left\{ e^{- i\theta}\left[ \sqrt {2\kappa _2} \, c(t-\tau ) - c_{\rm in}^{(2)}(t - \tau ) \right] \right. \nonumber \\ & +\left. e^{i\theta}\left[ \sqrt {2\kappa _2} \,{c^\dag }(t-\tau ) - c_{\rm in}^{(2)\dag}(t-\tau) \right] \right\}\nonumber \\&+ \sqrt {1-\eta } \, X_\nu (t-\tau )\,.
	\end{align}
	In the regime where pump detuning is much larger than the optomechanical coupling and the cavity decay rate ($ \Delta\gg \Lambda, \kappa $), it is convenient to rewrite the cavity delayed time operators as a product of  a slowly varying term $ \bar c_{{\rm{out}}}^{(2)}(t ) $ and a fast oscillating one as $ c_{{\rm{out}}}^{(2)}\left( {t - \tau } \right) \simeq \bar c_{{\rm{out}}}^{(2)}(t - \tau ){e^{ - i\Delta (t - \tau )}} $. If the feedback delay time is much shorter than both the characteristic time of the optomechanical  interaction and the decay time of the cavity, then the delay time dependence of the slowly varying part, $ \bar c_{{\rm{out}}}^{(2)}(t - \tau ), $ can be ignored, therefore we have $\bar c_{{\rm{out}}}^{(2)}(t - \tau )\simeq \bar c_{{\rm{out}}}^{(2)}(t )$. We thus can rewrite the output operator as
	\begin{equation}
		c_{{\rm{out}}}^{(2)}\left( {t - \tau } \right) \simeq \bar c_{{\rm{out}}}^{(2)}(t){e^{i\Delta t}}{e^{ - i\Delta \tau }} = c_{{\rm{out}}}^{(2)}(t){e^{-i\Delta \tau }}
	\end{equation}
	By defining a global phase $ \phi=\theta+\Delta \tau $, we then approximate Eq.~(\ref{photocurrent2}) as 
	\begin{equation}\label{photocurrent3}
	\begin{split}
		{i_{{\rm{fb}}}}(t - \tau ) =& \sqrt {\eta/2}  \left\{ {{e^{ - i\phi }}\left[ {\sqrt {2{\kappa _2}} c(t) - c_{{\rm{in}}}^{(2)}(t)} \right]} \right. \\
		&+ \left. {{e^{i\phi }}\left[ {\sqrt {2{\kappa _2}} {c^\dag }(t) - c_{{\rm{in}}}^{(2)\dag }(t)} \right]} \right\} + \sqrt {1 - \eta }  {X_\nu }(t)\,.
		\end{split}
	\end{equation}
	By using equations (\ref{input_feedback_noise}) and (\ref{photocurrent3}), we can rewrite the equation of motion for the cavity field as
	\begin{equation}\label{Equation_of_motion_2}
		\begin{split}
			\dot c =& -  \kappa \, c + i\Lambda b  + g\,\sqrt {2 \eta\kappa _1\kappa _2 } \left[ e^{ - i \phi}c(t ) + e^{i \phi}c^\dag (t ) \right]\\ &+ \sqrt {2\kappa _{\rm fb}}\, c_{\rm in,fb}(t), 
		\end{split}
	\end{equation}
	where we have used the feedback-modified cavity decay rate 
	\begin{equation}\label{feedback_modified_cavity_decay}
		\kappa_{\rm fb} = \kappa - g\sqrt {2\eta\kappa _1\kappa _2}\,,
	\end{equation}
	and the corresponding noise operator
	\begin{equation}
		\begin{split}
			c_{\rm in,fb}&(t) = \frac{1}{\sqrt {2\kappa_{\rm fb}}}\left\{ - g\sqrt {\kappa _1 \eta } \left[ {e^{-i\phi}c_{\rm in}^{(2)}(t) + e^{i\phi }c_{\rm in}^{(2)\dag }(t)}\right]  \right.\\
			&\left. + \sqrt {2\kappa_2} c_{\rm in}^{(2)}(t)  +g\sqrt {2\kappa _1 (1-\eta)} X_\nu(t) + \sqrt {2\kappa _1} c_{\rm in ,0}^{(1)}(t)
			\right\}\,.
		\end{split}
	\end{equation}
	The feedback-modified  noise operator obeys the following correlation functions
	\begin{align}
		\langle c_{\rm in,fb}(t) c_{\rm in,fb}(t') \rangle  =& \left(n_{\rm fb} - m_{\rm fb}\right) \delta (t - t')\,, \label{Optical_correlation1}\\
		\langle {c_{\rm in,fb}^\dag (t)c_{\rm in,fb}^\dag (t')} \rangle  =& \left(n_{\rm fb} - m_{\rm fb}^*\right)\delta (t-t') \,,\label{Optical_correlation2}\\
		\langle c_{\rm in,fb}^\dag (t)c_{\rm in,fb}(t') \rangle  =& n_{\rm fb}\delta (t - t')\,,\label{Optical_correlation3}\\
		\langle c_{\rm in,fb}(t)c_{\rm in,fb}^\dag (t') \rangle  =& \left( {{n_{{\rm{fb}}}}\! + {\kappa }/{{{\kappa _{{\rm{fb}}}}}} \!-  m_{\rm fb}\!- m_{\rm fb}^*} \right) \delta (t - t')\,, \label{Optical_correlation4}
	\end{align}
	where we have defined $ n_{\rm fb}  = \zeta ^2 /(4 \eta  \kappa_2 \kappa_{\rm fb}) $ as the feedback-mediated number of thermal excitations and $ {m_{{\rm{fb}}}} = \left( {\zeta /2{\kappa _{{\rm{fb}}}}} \right)\exp \left( {i\phi } \right) $, with feedback reduction of cavity linewidth $\zeta  = g\sqrt {2\eta\kappa _1\kappa _2}$. This reveals that the cavity decay rate, as well as the cavity frequency (see Eq.~(\ref{Equation_of_motion_2})), can be controlled by manipulating the feedback gain $g$ and the global phase $\phi$. Moreover, the system can be viewed as an engineered reservoir interacting with the cavity and mechanical mode.  Together with correlations (\ref{Mechanical_correlation1}-\ref{Mechanical_correlation2}) and (\ref{Optical_correlation1}-\ref{Optical_correlation4}) the quantum Langevin equations (\ref{equations_of_motion1}) and (\ref{Equation_of_motion_2}) describe the evolution of the cavity field and the mechanical motion of the nanomechanical membrane, including a feedback-controlled in-loop light and all fluctuation effects. We then summarize the equations of motion in the following compact matrix form
	\begin{equation}\label{equations_of_motion}
		\frac{{{\rm d}{\bf{u}}(t)}}{{{\rm d}t}} = {\bf{Au}}(t) + {\bf{H}}{{\bf{n}}_{{\rm{in,fb}}}}(t)  \,.
	\end{equation}
	Here, we have defined the vector of fields $  {\bf{u}}(t)=[c(t),c^\dagger(t),b(t),b^\dagger(t)]^T$, matrix ${\bf{H}} = {\rm{diag}}\left[ {\sqrt {2\kappa_{fb} } ,\sqrt {2\kappa_{fb} } ,\sqrt {2\gamma } ,\sqrt {2\gamma }  } \right] $ and the corresponding vector of noises ${{\bf{n}}_{{\rm{in,fb}}}}(t) = [ {{c_{{\rm{in}},{\rm{fb}}}}(t),c_{{\rm{in}},{\rm{fb}}}^\dag (t),{b_{{\rm{in,f}}}}(t),b_{{\rm{in,f}}}^\dag (t)} ]^T $. We have also defined the drift matrix $ {\bf A} $ as
	\begin{equation}\label{Drift_matrix}
		{\bf{A}} = \left( {\begin{array}{*{20}{c}}
				{\zeta {e^{ - i\phi }}   - \kappa }&{\zeta {e^{i\phi }}}&{i\Lambda }&{0 }\\
				{\zeta {e^{ - i\phi }}}&{\zeta {e^{i\phi }}   - \kappa }&{ 0 }&{ - i\Lambda }\\
				{i\Lambda }&{0 }&{ - \gamma }&0\\
				{ 0 }&{ - i\Lambda }&0&{- \gamma }
		\end{array}} \right)\,.
	\end{equation}
\subsection{System response in the frequency domain}
	Now that we understand the system dynamics, we can investigate how the feedback loop can be used for force sensing. For a detailed treatment and a review of the basic statistical properties of quantum noise, including its detection and a basic introduction to weak continuous measurements, see \cite{Clerk2010}. 
	Equation (\ref{equations_of_motion}) can be solved in the frequency space by taking the Fourier transform of all the operators and noise sources introduced via $ o[ \omega] = \smallint_{-\infty}^{+ \infty} {dt{e^{i\omega t}}o(t)} $. We then write the solution of Eq.~(\ref{equations_of_motion}) in the frequency domain as 
	\begin{equation}\label{Equations_of_motion_matrix_form1}
		{\bf{u}}[\omega ] =  - {\left( {{\bf{A}} + i\omega {{\bf{I}}_{4 \times 4}}} \right)^{ - 1}}{\bf{H}}{{\bf{n}}_{{\rm{in,fb}}}}[\omega ]  \,,
	\end{equation}
	where ${{\bf{I}}_{4 \times 4}}$ is the identity matrix.
	
	By substituting $ {\bf u}\left[ \omega  \right] $ into the input-output relation $ {{\bf{n}}_{{\rm{out}}}^{(2)}}[\omega ] =   {\bf{Gu}}[\omega ] - {{\bf{n}}_{{\rm{in}}}^{(2)}}[\omega ] $, which relates output and input fields, the fluctuations of the output fields are obtained as
	\begin{align}
		{{\bf{n}}_{{\rm{out}}}^{(2)}}[\omega ] &= -\left[ {{{{\bf{G}}\left( {{\bf{A}} + i\omega {{\bf{I}}_{4 \times 4}}} \right)}^{ - 1}}{\bf{H}} } \right]{{\bf{n}}_{{\rm{in,fb}}}}[\omega ]-{{\bf{n}}_{{\rm{in}}}^{(2)}}[\omega ]\nonumber\\
		&= {\bf{s}}[\omega ]{{\bf{n}}_{{\rm{in,fb}}}}[\omega ]-{{\bf{n}}_{{\rm{in}}}^{(2)}}[\omega ]\,,
	\end{align}
	where have defined $ {\bf{n}}_{{\rm{in}}}^{(2)} = {[c_{{\rm{in}}}^{(2)},c_{{\rm{in}}}^{(2)\dag },{b_{{\rm{in,f}}}},b_{{\rm{in,f}}}^\dag ]^T} $, $ {\bf{n}}_{{\rm{out}}}^{(2)} = {[c_{{\rm{out}}}^{(2)},c_{{\rm{out}}}^{(2)\dag },{b_{{\rm{out}}}},b_{{\rm{out}}}^\dag ]^T} $ and $ {\bf{G}} = {\rm{diag}}\left[ {\sqrt {2{\kappa _2}} ,\sqrt {2{\kappa _2}} ,\sqrt {2\gamma } ,\sqrt {2\gamma } } \right] $. 
	We have also introduced the scattering matrix $ {\bf{s}}[\omega ] $ as
	\begin{equation}\label{sactterin_matrix}
		{\bf{s}}[\omega ] =  - {\bf{G}}{\left( {{\bf{A}} + i\omega {{\bf{I}}_{4 \times 4}}} \right)^{ - 1}}{\bf{H}},
	\end{equation}
	Note that we use the convention $ {o^\dag }[\omega ] = {\smallint}_{- \infty }^{+\infty }dt e^{i\omega t} o^\dag (t) $ which implies $ o^\dag [\omega ] = (o[-\omega])^\dag$. Full expressions of the scattering matrix elements are given in appendix \ref{Appendix:B}. 
\section{	\label{sec:Sec3} Force measurement}
In this section, we describe force sensing beyond the SQL in the presence of the feedback loop. We start by deriving expressions for the mechanical response and the optically added noise. Then we show that the feedback loop can enhance the mechanical response and improve its bandwidth. We also show how noise suppression and signal amplification can be achieved by tuning the system parameters, particularly the asymmetry in the decay rates of the two mirrors. Moreover, the optically added noise and the force sensitivity will not change much compared to an open-loop system.
	\subsection{Spectral characteristics of force}
	An external force acting on the nanomechanical membrane displaces it from the equilibrium point which, via the optomechanical coupling, affects the optical cavity output.  As a consequence, the signal associated with the external force can be extracted by measuring the optical output generalized in-loop quadrature, $ X_{\rm out}^{\phi,(2)}  $, given in terms of the input fields by
	\begin{equation}\label{Generalized_quadrature}
		X_{{\rm{out}}}^{\phi ,(2)} = \frac{1}{{\sqrt 2 }}\left\{ {\Psi [\omega ]{c_{{\rm{in}},{\rm{fb}}}} + \Phi [\omega ]{b_{{\rm{in}},{\rm{f}}}} + {\rm{H}}{\rm{.c}}{\rm{.}}} \right\} - X_{{\rm{in}}}^{\phi ,(2)}{\mkern 1mu} ,
	\end{equation}
	where we have defined the functions $  \Psi [\omega ] $ and $  \Phi [\omega ]  $ in terms of the scattering matrix elements given in appendix \ref{Appendix:B} as
	\begin{align}\label{Psi}
		\Psi [\omega ] =& {s_{11}}{e^{ - i\phi }} + {s_{21}}{e^{i\phi }} \nonumber\\
		=& \frac{{2{e^{ - i\phi }}(\gamma  - i\omega )\sqrt {{\kappa _2}{\kappa _{{\rm{fb}}}}} }}{{ - 2\zeta (\gamma  - i\omega )\cos \phi  + (\gamma  - i\omega )(\kappa  - i\omega ) + {\Lambda ^2}}}\,,
	\end{align}
	\begin{align}\label{Phi}
		\Phi [\omega ] =& {s_{13}}{e^{ - i\phi }} + {s_{23}}{e^{i\phi }} \nonumber \\
		=& \frac{{2i\Lambda {e^{ - i\phi }}\sqrt {\gamma {\kappa _{\rm{2}}}} }}{{ - 2\zeta (\gamma  - i\omega )\cos \phi  + (\gamma  - i\omega )(\kappa  - i\omega ) + {\Lambda ^2}}}\,.
	\end{align}
 	We then use the output quadrature to estimate the applied external force on the nanomechanical membrane. Using the expressions for the elements of the scattering matrix, one finally obtains the following result for the transmitted optical quadrature:
	\begin{equation}
		X_{{\rm{out}}}^{\phi,(2)}  = {\left. {X_{{\rm{out}}}^{\phi,(2)} } \right|_{{F } = 0}} + X_F^\phi \,,
	\end{equation}
	where the contribution of the quantum noise $  {X_{{\rm{out}}}^{\phi,(2)}} |_{F = 0} $ can be obtained by replacing 
	$ {b_{{\rm{in,f}}}} \to {b_{{\rm{in}}}} $ and $ b_{{\rm{in,f}}}^\dag  \to b_{{\rm{in}}}^\dag  $ in Eq.~(\ref{Generalized_quadrature}). 
	Moreover, the contribution of the external force is 
	\begin{align} 
		X_F^\phi  =&  - \frac{i}{{2\sqrt {2 \hbar m\gamma {\omega _{\rm{m}}}} }}\left\{ ({s_{13}}{e^{ - i\phi }} + {s_{23}}{e^{i\phi }})F[\omega  + {\omega _{\rm{m}}}] \right.\nonumber\\ & \left.- ({s_{14}}{e^{ - i\phi }} + {s_{24}}{e^{i\phi }})F[\omega  - {\omega _{\rm{m}}}] \right\} \nonumber\\
		 =& \frac{i}{{\sqrt {2 \hbar m\gamma {\omega _{\rm{m}}}} }}\bar F[\omega ]\, ,
	\end{align}
	where we define the transduction force by the force estimator
	\begin{equation}\label{Force_estimator}
		\bar F[\omega ] = \frac{1}{2}\left( {{\Phi ^*}[\omega ]F[\omega  - {\omega _{\rm{m}}}] - \Phi [\omega ]F[\omega  + {\omega _{\rm{m}}}]} \right)\,.
	\end{equation} 
	We then define the effective force operator as \cite{Xu}
	\begin{equation}\label{force operator}
		 F_{\rm eff}[\omega ] = \frac{X_{\rm out}^{\phi,(2)} [\omega ]}{\partial X_{\rm out}^{\phi,(2)} [\omega ]/\partial \bar F [\omega ]} = N^\phi[\omega ] + \bar F [\omega ]\,,
	\end{equation}
	where the noise force operator is given by
	\begin{equation}
		{N^\phi }[\omega ] = i\sqrt {2\hbar m\gamma {\omega _{\rm{m}}}} {\rm{ }}{\left. X_{{\rm{out}}^{\phi,(2)}} \right|_{{F} = 0}}.
	\end{equation}
	We then consider the spectral density of the noise force as
	\begin{align}\label{Eq:spectrum}
		\delta (\omega  + \omega ')S_{NN}^\phi (\omega ) = \frac{1}{{4\pi }}\langle {N^\phi }[\omega ]{N^{\phi,\dagger} }[\omega '] + (\omega  \leftrightarrow \omega ')\rangle\,,
	\end{align}
	where the notation $ (\omega  \leftrightarrow \omega ') $ denotes the swapping of the arguments of the first term. 
	After straightforward calculations, the spectrum of the noise force operator, $ {N^\phi } $, is given by
	\begin{equation} \label{Spectrum_of_the_noise_force}
		S_{NN}^\phi (\omega ) =2 \hbar m\gamma {\omega _{\rm{m}}} R_{\rm m}[\omega] \left[ {\left( {\bar n + \frac{1}{2}} \right) + {n_{{\rm{add}}}}[\omega ]} \right]\,,
	\end{equation}
	where we have used correlations in the frequency domain see appendix \ref{Appendix:C} and  $ {R_{\rm{m}}}[\omega ] $ is the mechanical response defined by
	\begin{equation}\label{mechanical_response}
		{R_{\rm{m}}}[\omega ] =\frac{1}{2} ( \Phi [ - \omega ]{\Phi ^*}[\omega ] + \Phi [\omega ]{\Phi ^*}[ - \omega ])\,.
	\end{equation}
	 In Eq.~(\ref{Spectrum_of_the_noise_force}), ${n_{{\rm{add}}}}[\omega ]$ is the noise of the measurement process which can be interpreted as an effective increase in the number of thermal excitations of the mechanical reservoir due to the backaction of the optical mode. After evaluating lengthy but straightforward calculations, the added measurement noise is given by
	\begin{align}\label{added_noise}
		{n_{{\rm{add}}}}[\omega ] =& \frac{1}{{{2 R_{\rm{m}}}[\omega ]}}\Big\{ 
		\big[2\Psi [\omega ]\Psi [ - \omega ]\left( {{n_{{\rm{fb}}}} - {m_{{\rm{fb}}}}} \right) + {\rm c.c. }\big]   \big. \nonumber\\
		& + \big. \big(\Psi  [\omega  ]{\Psi ^*}[ - \omega ]+  {\rm{c}}{\rm{.c.}} \big)\big(2{n_{{\rm{fb}}}} + \frac{\kappa }{{{\kappa _{{\rm{fb}}}}}} - {m_{{\rm{fb}}}} - m_{{\rm{fb}}}^*\big) \big. \nonumber\\
		& \big.+	2 - \left( {\Psi [\omega ] + \Psi [ - \omega ] + {\rm{c}}{\rm{.c}}} \right)\left( {{e^{i\phi }}p_{{\rm{fb}}}^*[\omega ] + {\rm c.c.}} \right)  \big. \nonumber\\
		&\big. - \left[ {{e^{i\phi }}\left( {\Psi [ - \omega ] + \Psi [\omega ]} \right) + {\rm{c}}{\rm{.c}}{\rm{.}}} \right]\sqrt {{\kappa _{\rm{2}}}/{\kappa _{{\rm{fb}}}}} 	\Big\},
	\end{align} 
	with $ 	{p_{{\rm{fb}}}}= (-\zeta/\sqrt {4{\kappa _2}{\kappa _{{\rm{fb}}}}}) \exp (i\phi) $.
	
 	The transduction of the external force and the mechanical inputs onto the light field is characterized by Eq.~(\ref{Generalized_quadrature}). This transduction can be obtained in terms of the corresponding scattering matrix elements given in Appendix \ref{Appendix:B} and finally translated into functions $ \Phi[\omega] $ and $ \Phi^*[\omega] $. In the case of an ideal resonant impulsive force, $ F[\omega]=\textrm{const.} $, the force estimator $ \bar F [\omega ] $ becomes
	\begin{equation}\label{key}
		\bar F [\omega ]= {R_{\rm{m}}}[\omega ] \cos^2 \phi F\,.
	\end{equation}	
\subsection{Mechanical response and the added noise}
	We next calculate the added noise and the mechanical response given by Eqs.~(\ref{added_noise}) and (\ref{mechanical_response}), respectively, in the resolved sideband regime in the limit where the mechanical frequency is much	larger than the cavity decay rate, $ {\omega _{\rm{m}}} \gg \kappa  $. The frequency-dependent mechanical response and the optically added noise in the rotating wave approximation (RWA) (with calculations beyond the RWA delegated to the appendix \ref{Appendix:A}) are represented by
	\begin{equation}\label{Mechanical_Response}
		{R_{\rm{m}}}[\omega ] = \frac{4 {\cal C} \gamma ^2 \kappa  \kappa _2}{|({\cal C}+1) \gamma  \kappa -2 \zeta  (\gamma -i \omega ) \cos \phi -i \omega  (\gamma +\kappa )-\omega ^2|^2 }\,,
	\end{equation}
	\begin{align}\label{Added_noise}
		{n_{{\rm{add}}}}\![\omega  ]\!=&\!\frac{\omega ^2 \! \left(\gamma ^2\!-2 {\cal C} \gamma  \kappa \!+\kappa ^2\right)\!+\!({\cal C}\!+1\!)^2 \!\gamma ^2 \kappa ^2\! -\! 4 {\cal C} \gamma ^2 \kappa  \kappa _2+\omega ^4}{4 {\cal C} \gamma ^2 \kappa  \kappa _2} \nonumber\\
		& + \frac{{\left( {{\omega ^2} + {\gamma ^2}} \right)(1 - \eta ){\zeta ^2}}}{{2 {\cal C}{\gamma ^2}\eta \kappa {\kappa _2}}}\,,
	\end{align}
	where $ {\cal C}= \Lambda^2/\kappa \gamma $ is the optomechanical cooperativity. Eqs. (\ref{Mechanical_Response}), (\ref{Added_noise}) are expressed in the interaction picture, where $ \omega = 0 $ means that we look at detection at the cavity resonance frequency. In this case, we are looking at forces that are quasi-resonant with the mechanical resonator. The on-resonance mechanical response and the optically added noise are given by
	\begin{align} 
		R_{\rm{m}}^{{\rm{fb}}}[\omega  = 0] &=	\frac{4 {\cal C} \kappa  \kappa _2}{({\cal C} \kappa -2 \zeta  \cos \phi +\kappa )^2}\,, \label{mechanical_response_resonance}\\
		n_{{\rm{add}}}^{{\rm{fb}}}[\omega  = 0] &= \frac{{{{\left( {{\cal C} - 1} \right)}^2}}}{{4{\cal C}}} +\frac{({\cal C}+1)^2 \kappa_1 }{4 {\cal C} \kappa _2}+\frac{{(1 - \eta ){\zeta ^2}}}{{2{\cal C}\eta \kappa {\kappa _2}}}\, . \label{added_noise_resonance}
	\end{align}
	 A remarkable result from Eq.~(\ref{mechanical_response_resonance}) is that the on-resonance mechanical response is never larger than unity for an open loop system. That is, a single phonon at the mechanical input can be converted to only one photon at the output which is a limitation in the functionality of the open-loop system in the red detuned regime. Therefore according to Eq.~(\ref{mechanical_response_resonance}), the open-loop system provides only direct transduction of the force and any force-detection gain is impossible. 
	 
	We next turn to the numerical calculations of the added noise and mechanical response by using parameters feasible in a typical experimental setup of membrane-in-the-middle cavity optomechanical systems \cite{Rossi,Rossi2}. We consider a mode of vibration with an effective mass of $m= 10^{-12}\, \rm kg$ inside an optical cavity with decay rate $\kappa=\kappa_1+\kappa_2 = 0.06\omega_{\rm m }$. The mechanical decay rate is $\gamma/\omega_{\rm m} =  3.4 \times {10^{ - 6}}$ and the mechanical frequency is $\omega_{\rm m}=2\pi  \times 343.13\,{\rm{kHz}}$. The optomechanical coupling  rate, the local oscillator phase, the cavity detuning and the feedback gain could be tuned in situ. 

	\begin{figure}
		\includegraphics[width=8.6cm]{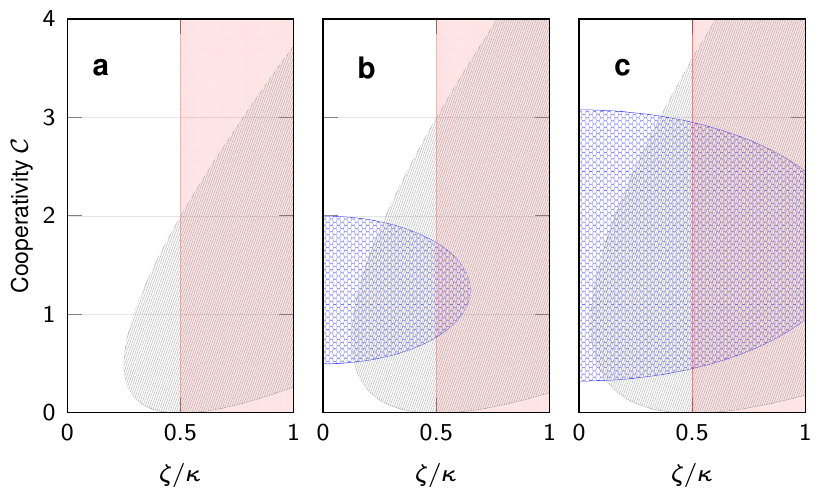}
		\caption{The regions of standard quantum limit suppression $ n_{\rm add}[0]<1/2 $ (blue honeycomb pattern), signal amplification $ R_{\rm m}[0]>1 $ (gray hatched pattern) and the instability (red solid area) as functions of the optomechanical cooperativity and feedback gain for detection efficiency $ \eta=0.6 $.  Symmetric cavity with $\kappa_2/\kappa= 0.5$ (a) and asymmetric cavity with  $ \kappa_2/\kappa=0.75 $ (b) and $ \kappa_2/\kappa=0.90 $ (c).  We consider an optical cavity with decay rate $\kappa=\kappa_1+\kappa_2 = 0.06\omega_{\rm m }$, the mechanical decay rate  $\gamma/\omega_{\rm m} =  3.4 \times {10^{ - 6}}$ and the mechanical frequency  $\omega_{\rm m}=2\pi  \times 343.13\,{\rm{kHz}}$. }
		\label{Fig:Fig2}
	\end{figure}
	In Fig.~\ref{Fig:Fig2}, we show the regions of SQL suppression $ n_{\rm add} [0] <1/2$ \cite{Clerk2010,Hertzberg,Schreppler} (blue honeycomb pattern) and signal amplification $ R_{\rm m}[0]>1 $ (gray-hatched pattern) versus optomechanical cooperativity and feedback gain for three different cavity configurations. The denominator of the mechanical response given by Eq.~(\ref{mechanical_response_resonance}) must tend to its smallest possible value to achieve signal amplification.  This puts a constraint on the angle $- \pi /2 < \phi  < \pi /2 $. Hereafter, we take the global phase $\phi$ such that $\cos\phi = 1$ to have maximum amplification. Signal amplification starts when we set $\zeta\simeq \kappa/4$ and $ {\cal C}=0.5$ for a symmetric cavity configuration. As is evident from Fig.~\ref{Fig:Fig2}, the sub-SQL region increases as we increase the ratio of $\kappa_2/\kappa$. For an asymmetric configuration, $ \kappa_2 > \kappa_1$, the regions are much wider than for a symmetric configuration. Generally, simultaneous noise suppression and signal amplification can not be achieved for relatively small values of optomechanical cooperativity and feedback gain. What is noticeable is that the feedback plays its role in signal amplification in the red detuning regime. The additional requirement for performing the force measurement is that the system under consideration should be stable. Here, the unstable region is shown by a red solid region (see appendix \ref{Appendix:E} for details on the system's stability conditions). 
	
	\begin{figure}
		\includegraphics[width=8.6cm]{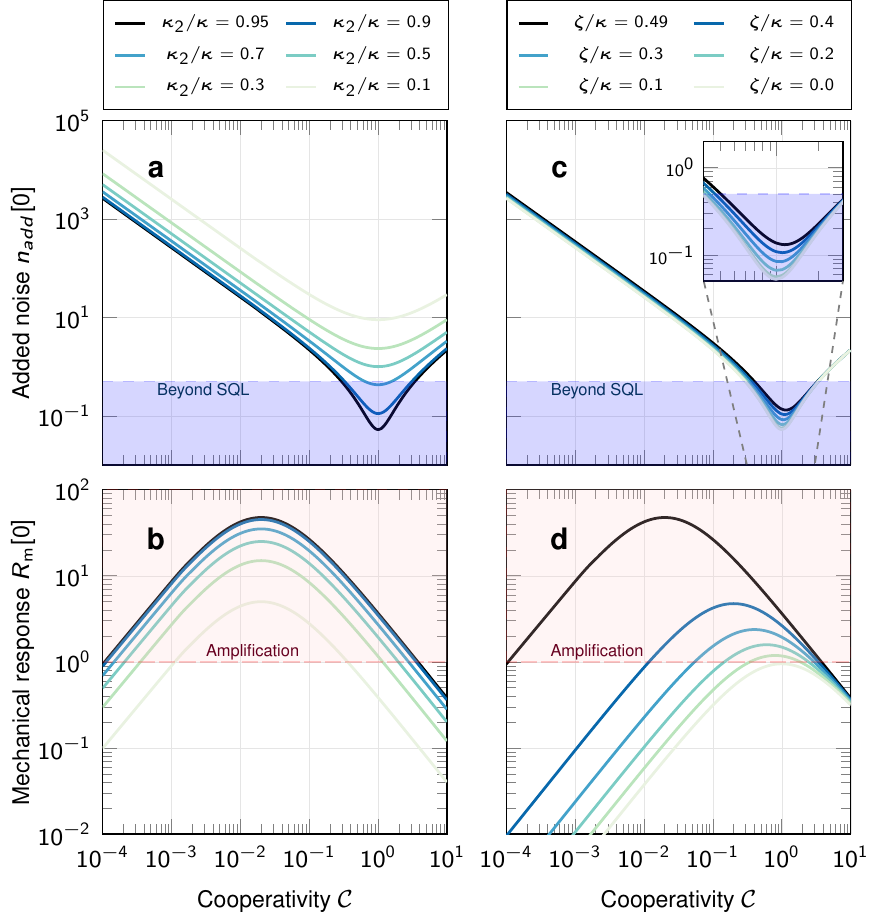}
		\caption{(a) The on-resonant optically added noise of measurement and (b) the mechanical response as a function of the optomechanical cooperativity for different ratios of  $\kappa_2/ \kappa$, fixed values of the feedback gain $ \zeta/\kappa=0.49 $ and the detection efficiency $ \eta=1 $. (c) The on-resonant optically added noise and (d) the mechanical response as a function of the optomechanical cooperativity for different values of feedback gain, fixed  $ \kappa_2/\kappa=0.95 $ and detection efficiency $ \eta=0.6 $. The line with $\zeta/\kappa=0 $ corresponds to the no feedback scenario. Other parameters given in caption of Fig. \ref{Fig:Fig2}. The noise suppression (beyond SQL, $ n_{\rm add}[0]<1/2 $) and signal amplification ($ R_{\rm m}[0]>1 $) regions are shown in blue and red, respectively. }	
		\label{Fig:Fig3}
	\end{figure}

Based on Eqs.~(\ref{mechanical_response_resonance}) and (\ref{added_noise_resonance}), the decay rates of the cavity mirrors influence the system dynamics.
We drive the system through the mirror ${\bf M}_1$ which is highly reflective since we want to reduce any optical losses on this cavity end. This high reflectivity then requires strong drive and high feedback gain but, at the same time, ensures that the added noise remains low. On the other hand, the signal we detect has to be sufficiently large so the mirror ${\bf M}_2$ has to be more transmissive. Indeed, in the symmetric case, $ \kappa_1=\kappa_2 $,  it is clear that in the limit of ${\cal C} \rightarrow 1 $, the added optical noise in Eq.~(\ref{added_noise_resonance}) is $ {n_{{\rm{add}}}}[\omega  = 0] =1 $, and  the maximum value of the mechanical response is $ R_{\rm{m}}^{{\rm{fb}}}[\omega  = 0]=2$ (assuming unit detection efficiency). The minimum added noise is obtained by using an asymmetric cavity with the mirror ${\bf M}_1 $ \emph{less transmissive} than the mirror ${\bf M}_2$. Therefore, this proposal will be optimal in situations where we use an asymmetric cavity $ \kappa_1<\kappa_2 $. In the strongly asymmetric case, $ \kappa_1\ll \kappa_2 $, with unit detection efficiency, in the limit of ${\cal C} \rightarrow 1 $, the added optical noise is, $ {n_{{\rm{add}}}}[\omega  = 0] \simeq {\kappa _1}/{\kappa _2} $, and the maximum achievable  mechanical response is $ R_{\rm{m}}^{{\rm{fb}}}[\omega  = 0]=4$. 

The on-resonance optically added noise and the mechanical response versus optomechanical cooperativity for different cavity asymmetries and a fixed value of feedback gain are shown in Figs.~\ref{Fig:Fig3} (a) and (b). We see that asymmetric cavities with $ \kappa_2>\kappa_1 $ are preferable for suppressing noise below SQL and amplifying the signal. Although the backaction and the feedback loop introduce measurement noise to the force measurement, with an appropriate choice of parameters, the measurement noise can be significantly reduced below the SQL. For instance, by using a high--efficiency detector ($\eta \simeq 1$), the last term in Eq.~(\ref{added_noise_resonance}), which is due to the feedback loop, can be neglected. In Figs.~\ref{Fig:Fig3} (c) and (d), we show the optically added noise and the mechanical response as a function of the optomechanical cooperativity for different values of feedback gain and fixed value of the cavity decay ratio. We have a large gain for small cooperativities at higher feedback gain, but the cost of the strong signal amplification is an increase in the added measurement noise. Nevertheless, the added noise is not excessively large in the region of interest (overlap of the amplification, sub-SQL and stable regions) to influence a precise force measurement. 

	\begin{figure}
		\includegraphics[width=8.6cm]{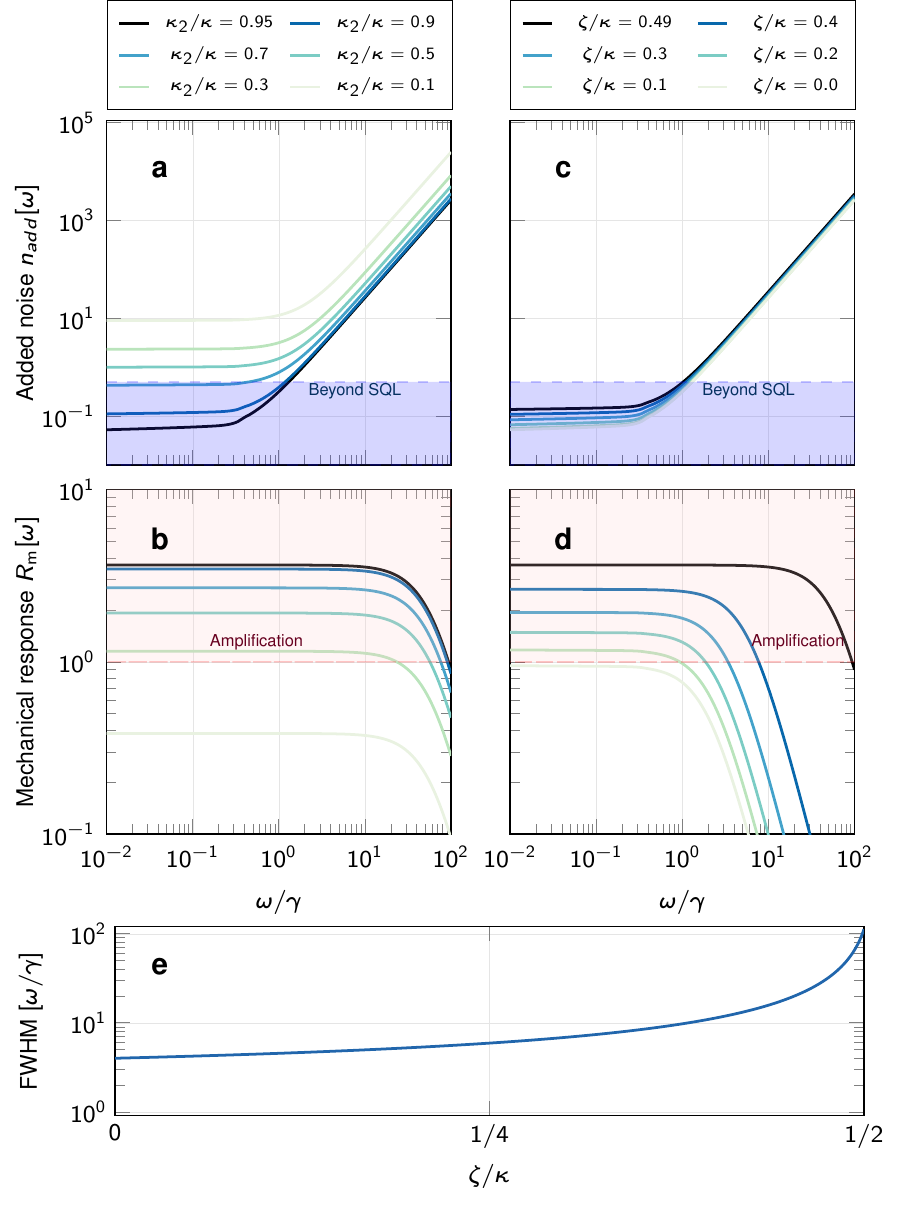}
		\caption{(a) The off-resonant optically added noise of measurement and (b) the mechanical response as a function of the  normalized frequency for different ratios of $\kappa_2/ \kappa$ for fixed values of the feedback gain $ \zeta/\kappa=0.49 $ and the optomechanical cooperativity ${\cal C}=1$.  (c) The off-resonant optically added noise of measurement and (d) the mechanical response as a function of the normalized frequency for different values of feedback gain and fixed  $ \kappa_2/\kappa=0.95 $, optomechanical cooperativity ${\cal C}=1$ and detection efficiency $ \eta=0.6 $. The line with $\zeta/\kappa=0 $ corresponds to the no feedback scenario. The noise suppression (beyond SQL, $ n_{\rm add}<1/2 $) and signal amplification ($ R_{\rm m}>1 $) regions are shown by blue and red area, respectively. (e) FWHM as a function of the feedback gain. Other parameters given in the caption of Fig. \ref{Fig:Fig2}.}	
		\label{Fig:Fig4}
	\end{figure}
	The optically added noise and the mechanical response away from the resonance ($ \omega \ne 0 $) for different ratios of $\kappa_2/ \kappa$ and for a fixed value of feedback gain  and optomechanical cooperativity ${\cal C}=1$ are depicted in Figs.~\ref{Fig:Fig4} (a) and (b). Again it helps to have a largely asymmetric cavity and high feedback gain to achieve strong mechanical response. In Figs.~\ref{Fig:Fig4} (c) and (d), we plot the off-resonant optically added noise and the mechanical response as a function of the normalized frequency. One can see that the range of signal frequencies over which there is amplification increase as we increase the feedback gain. Outside a certain frequency band, the mechanical response to the input signal drops compared to the response at resonance. We quantify the bandwidth of the amplification by the full width at half maximum (FWHM) of the mechanical response. That is, we solve the equation $R_{\rm m}[\omega]=R_{\rm m}[0]/2$ and the difference between the two values of the frequencies at which the mechanical response is equal to half of its maximum value is chosen to be FWHM. In Fig.~\ref{Fig:Fig4} (e), we plot FWHM as a function of feedback gain to show the increase in amplification bandwidth. This way, the total detection bandwidth is improved with high gain since the spectrum of the added noise is almost unaffected by the feedback. Moreover, there is no limitation on the gain-bandwidth product. That is, as we increase the mechanical response, we also increase the amplification bandwidth. 
\subsection{Force sensitivity}
	We use the homodyne photocurrent of Eq.~(\ref{Generalized_quadrature}) to perform force measurements \cite{Vitali2002}. In the case of stationary spectral measurements the signal $ S(\omega) $ is
	\begin{align}\label{signal}
		{\cal S} [ \omega  ] = \left| {\left\langle {X_{{\rm{out}}}^{\phi,(2)} [\omega ]} \right\rangle } \right|\,.
	\end{align}
	The noise associated with the signal $ S\left( \omega  \right) $ is given by its standard deviation
	\begin{align}\label{noise}
		{\cal N}[\omega ] = {\left[ {\frac{1}{2}{{\left\langle {\left[ {X_{{\rm{out}}}^{\phi,(2)} \left( \omega  \right)X_{{\rm{out}}}^{\phi,(2)} \left( { - \omega } \right) + \left( {\omega  \leftrightarrow  - \omega } \right)} \right]} \right\rangle }_{F = 0}}} \right]^{1/2}}\, ,
	\end{align}
	and it is evaluated in the absence of the external force $ F = 0 $. The signal-to-noise ratio (SNR) is defined to be the ratio of the signal, Eq.~(\ref{signal}), to the variance of all the noises, Eq.~(\ref{noise}), of a given measured signal and it quantifies how much the signal is corrupted by the noise introduced by the system \cite{Bernal,Vitali2002}. Here, we are interested in spectral measurements that are stationary, that is, in the limit in which the measurement time approaches infinity \cite{Bernal,Vitali2002}. We define SNR as  
	\begin{align}\label{SNR}
		{\rm SNR}[\omega] &= \frac{{\cal S} [ \omega  ]}{{\cal N}[\omega ] }  = \frac{{\left| {\bar F[\omega ]} \right|}}{{\sqrt {2\hbar m\gamma {\omega _{\rm{m}}}} {{R_{\rm m}[\omega]\left( {\bar n + 1/2 + {n_{{\rm{add}}}}[\omega ]} \right)}^{1/2}}}} \nonumber\\
		&= \frac{{\left| {\left( {{\Phi ^*}[\omega ]F[\omega  - {\omega _{\rm{m}}}] - \Phi [\omega ]F[\omega  + {\omega _{\rm{m}}}]} \right)/2} \right|}}{{\sqrt {2\hbar m\gamma {\omega _{\rm{m}}}} R_{\rm m}[\omega] {{\left( {\bar n + 1/2 + {n_{{\rm{add}}}}[\omega ]} \right)}^{1/2}}}}\, ,
	\end{align}
	In the case of an ideal resonant impulsive force, $ F[\omega]=\textrm{const.} $, SNR of the system becomes
	\begin{equation} \label{SNR2}
		{\rm SNR}[\omega ] = \frac{{ \left| {\cos \phi } \right|F}}{{\sqrt {2 \hbar m\gamma {\omega _{\rm{m}}}} {{\left( {\bar n + 1/2 + {n_{{\rm{add}}}}[\omega ]}\right)}^{1/2}}}}\,.
	\end{equation}
As expected,  the SNR of the system depends on the optically added noise, the thermal noise and the global phase. Since the signal enters into the system through the same channel as the thermal noise, the mechanical gain amplifies both the input signal and the mechanical thermal and quantum noises. That is, the enhancement in the mechanical gain does not lead to the improvement of the force sensitivity and the SNR. Although the added noise and the thermal noise can be reduced for an open-loop system, the mechanical response never becomes greater than unity. In contrast, in closed-loop dynamics, we can simultaneously increase the mechanical response by the feedback loop and reduce the added noise by the red-detuned laser drive. 

\begin{figure}
	\includegraphics[width=8.6cm]{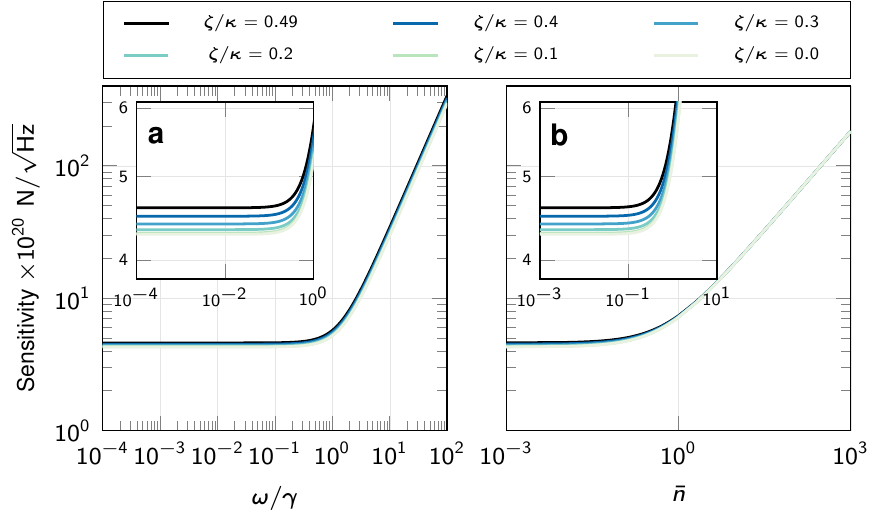}
	\caption{		
		(a) Force sensitivity at zero temperature as a function	of the normalized frequency and (b) force sensitivity at finite temperature for different values of feedback gain and fixed $ \kappa_2/\kappa=0.95 $, optomechanical cooperativity ${\cal C}=1$ and detection efficiency $\eta=0.6$. The line with $\zeta/\kappa=0 $ corresponds to the no-feedback scenario.  Other parameters given in the caption of Fig. \ref{Fig:Fig2}
}
	\label{Fig:Fig5}
\end{figure}

	We now turn to the investigation of the force sensitivity of the system. We quantify the sensitivity of the measurement by the system SNR \cite{Bernal}, given by Eq~(\ref{SNR2}). Performing a more sensitive measurement requires a greater associated SNR \cite{Bernal}. There is a minimum value of SNR that allows the detection of the force signal. The minimum detectable input of the device is the minimum signal required to produce an output with ${\rm SNR} [\omega]=1$ \cite{Lucamarini,Motazedifard,Xu,Bernal}. Therefore based on Eq~(\ref{SNR}), the sensitivity of a detector describes the minimum magnitude of the input signal that can be faithfully detected and measured, and it is defined as 
	\begin{equation}\label{sensitivity}
		{\rm S}[\omega] =\sqrt{2\hbar m\gamma {\omega _{\rm{m}}}} {\left[ {\left( {\bar n + \frac{1}{2}} \right) + {n_{{\rm{add}}}}[\omega ]} \right]^{1/2}}\,.
	\end{equation}
	In  Fig. \ref{Fig:Fig5} (a), we plot the force sensitivity as a function of the normalized frequency for different values of the feedback parameter and inefficent detection $ \eta=0.6 $. We see that although we have nearly the same sensitivity for all feedback parameters, inefficent detection reduces the force sensitivity.
	As can be deduced from Eq.~(\ref{sensitivity}), the less noise we have, the better the sensitivity we can achieve. As a result, Eq.~(\ref{sensitivity}) suggests that we can improve the sensitivity by setting the value of the added noise as small as possible. As shown in Fig.~\ref{Fig:Fig5} (b), thermal noise dramatically reduces the sensitivity of the system. Therefore, for a high-precision force sensing, we need to cool the environment to keep the thermal noise $\bar{n}$ as small as possible. For instance, for a system at room temperature $ T=300\,K $ ($ \bar n=1.8 \times 10^{7} $), the  value of sensitivity is  $ {\rm S}= 2.5\times 10^{-16} \,\,{\rm N}/\sqrt{\rm Hz} $, which can be improved  to $ {\rm S}= 4.1\times 10^{-20} \,\,{\rm N}/\sqrt{\rm Hz} $ by cooling the mechanical motion to zero temperature using cooling techniques. 
\section{\label{sec:Sec4} Summary and Conclusion}
We present a scheme for highly sensitive detection of forces achievable with state-of-the-art optomechanical systems \cite{Rossi,Rossi2,Mason2019}. The scheme manifests the potential of current experimental setups of membrane-in-the-middle cavity optomechanical systems to be employed as high-precision force sensors. The proposed force sensor is based on the optomechanical interaction between an optical mode and a nanomechanical membrane assisted by feedback-controlled in-loop light. We have shown that the optical cavity subjected to the feedback loop is a powerful tool for manipulating mechanical systems. An asymmetric optomechanical cavity in the resolved sideband regime ensures low measurement noise; the feedback loop provides a high mechanical gain and large detection bandwidth.

The applicability of our method depends strongly on the used parameter regime. In the classical regime, where thermal noise is the main limiting factor, amplifying the mechanical response using measurement feedback does not provide an advantage. As the SNR  and the sensitivity of the system are not improved compared to open-loop dynamics, the closed-loop system performs equivalently. In the regime of quantum sensors, where thermal noise is small ($\bar n$ is of order one or less), the situation is, however, completely different. First, if one decreases the measurement noise $ n_{\rm add} $, its effect will be more significant on a relative scale: one can go below the SQL by reducing the added noise to a negligible level. Second, it is also useful to amplify the mechanical motion to reduce the risk of the homodyne detector not registering any signal at all. With that one amplifies thermal noise as well, but as its negative effect is not significant for $ \bar n \simeq 1$, it is a price worth paying. In summary, in this regime, noise reduction and signal amplification can be both extremely useful, which is impossible to achieve in the open-loop case.

In the future, the proposed optomechanical setup could be modified to a single-sided cavity with an optical isolator. Such architecture would exhibit reduced optical loss, thus enhancing the effect of feedback on the system, and allow combination with squeezed vacuum injection for further increase of the measurement sensitivity. At the same time, such a system has one parameter less for optimization and would require an optical isolator to separate the output signal from input driving field. A detailed comparison of the performance of the two cavities could then provide important insight into the performance of feedback schemes in optomechanical systems.
	\begin{acknowledgments}
		We would like to thank Stefano Zippilli and Ali Motazedifard for their useful conversations. F. B. and L. R. acknowledge the support of project 19-22950Y of the Czech Science Foundation. O. \v{C}. and R. F. acknowledge the project of 19-17765S of Czech Science Foundation and the project CZ.02.1.01/0.0/0.0/16\_026/0008460 of MEYS CR. D. V. acknowledges the support of the European Union Horizon 2020 Programme for Research and Innovation through the Project QuaSeRT funded by the QuantERA ERA-NET Cofund in Quantum Technologies and the Project No. 862644 (FET Open QUARTET).
	\end{acknowledgments}
\appendix
\section{\label{Appendix:A} Beyond the rotating wave approximation}
	We investigate the system dynamics beyond RWA by using the Floquet approach \cite{Malz,Malz2}.  When including counter-rotating terms, the time-dependent equations of motion arise as 
	\begin{equation}\label{equations_of_motion_BRWA}
		\frac{{{\rm d}{\bf{u}}(t)}}{{{\rm d}t}} = {\bf{A}}(t){\bf{u}}(t) + {\bf{H}}{{\bf{n}}_{{\rm{in,fb}}}}(t)  \,.
	\end{equation}
	We can write $ {\bf{A}}(t) = \sum\nolimits_{n =  - 1}^1 {{{\bf{A}}^{(n)}}{e^{ - 2in{\omega _{\rm{m}}}t}}}  = {{\bf{A}}^{( - 1)}}{e^{ - 2i{\omega _{\rm{m}}}t}} + {{\bf{A}}^{(0)}} + {{\bf{A}}^{( 1 )}}{e^{2i{\omega _{\rm{m}}}t}}$ where $ {{\bf{A}}^{(0)}}={{\bf{A}}} $ in Eq.~(\ref{Drift_matrix}) and 
	\begin{equation} 
		{{\bf{A}}^{( 1 )}} = \left({\begin{array}{*{20}{c}}
			0&0&0&{i\Lambda }\\
			0&0&0&0\\
			0&{i\Lambda }&0&0\\
			0&0&0&0
		\end{array}} \right)\,, \\
		\quad {{\bf{A}}^{( -1 )}} = \left( {\begin{array}{*{20}{c}}
			0&0&0&0\\
			0&0&{ - i\Lambda }&0\\
			0&0&0&0\\
			{ - i\Lambda }&0&0&0
		\end{array}} \right)\,.
	\end{equation}
	By defining the system in terms of its Fourier components as 
	\begin{equation}
		{\bf{u}}(t) = \sum\nolimits_{n =  - \infty }^\infty {{{\bf{u}}^{(n)}}(t){e^{ - 2in{\omega _{\rm{m}}}t}}} \,,
	\end{equation}
	we can write Eq.~(\ref{equations_of_motion_BRWA}) in the frequency domain in the following matrix form
	\begin{align}\label{equations_of_motion_BRWA_total}
		\left( { - i\omega  - 2in{\omega _{\rm{m}}}} \right){{\bf{u}}^{(n)}}[\omega ]& = {{\bf{A}}^{( - 1)}}{{\bf{u}}^{(n - 1)}}[\omega ] + {{\bf{A}}^{(0)}}{{\bf{u}}^{(n)}}[\omega ] \nonumber \\ &+ {{\bf{A}}^{(1)}}{{\bf{u}}^{(n + 1)}}[\omega ] + {\delta _{n,0}}{\bf{H}}{{\bf{n}}_{{\rm{in}},{\rm{fb}}}}[\omega ].
	\end{align}
	In RWA, $ {{\bf{A}}^{( n \ne 0 )}}=0 $, we neglect terms containing $ \omega\pm 2\omega_{\rm{m}} $ in these equations.  Counter-rotating components introduce sidebands shifted by $\pm 2\omega _{\rm{m}} $. We can then write Eq.~(\ref{equations_of_motion_BRWA_total}) as
	\begin{equation} 
		{\bf{\bar u}}[\omega ] =  - {\left( {i\omega {{\bf{I}}_{12 \times 12}} + {\bf{\bar A}}} \right)^{ - 1}}{\bf{\bar H}}{{{\bf{\bar n}}}_{{\rm{in}},{\rm{fb}}}}[\omega ]\,,
	\end{equation}
	where ${{\bf{I}}_{12 \times 12}}$ is the identity matrix and we define the vector of fluctuation as $ \bar{\bf u}=\left[ {{{\bf{u}}^{( - 1)}}[\omega ],{{\bf{u}}^{(0)}}[\omega ],{{\bf{u}}^{(1)}}[\omega ]} \right] $ and its corresponding noise vector as $ {{{\bf{\bar n}}}_{{\rm{in}},{\rm{fb}}}}[\omega ] = \left[0,{{\bf{n}}_{{\rm{in}},{\rm{fb}}}}[\omega ],0 \right] $. Moreover, $ {\bf{\bar H}} = {\rm{diag}}[0,{\bf{H}},0] $ and the matrix ${\bf{\bar A}} $ is given by 
	\begin{equation}
		{\bf{\bar A}} =
	\left( {\begin{array}{*{20}{c}}
			{{{\bf{A}}^{(0)}} - 2i{\omega _{\rm{m}}}{{\bf{I}}_{4 \times 4}}}&{{{\bf{A}}^{(1)}}}&0\\
			{{{\bf{A}}^{( - 1)}}}&{{{\bf{A}}^{(0)}}}&{{{\bf{A}}^{(1)}}}\\
			0&{{{\bf{A}}^{( - 1)}}}&{{{\bf{A}}^{(0)}} + 2i{\omega _{\rm{m}}}{{\bf{I}}_{4 \times 4}}}
	\end{array}} \right)\,.
	\end{equation}
	By substituting $\bar {\bf u}\left[ \omega  \right] $ into the input-output relation the fluctuations of the output fields can be obtained. 

\begin{figure}
		\includegraphics[width=8.6cm]{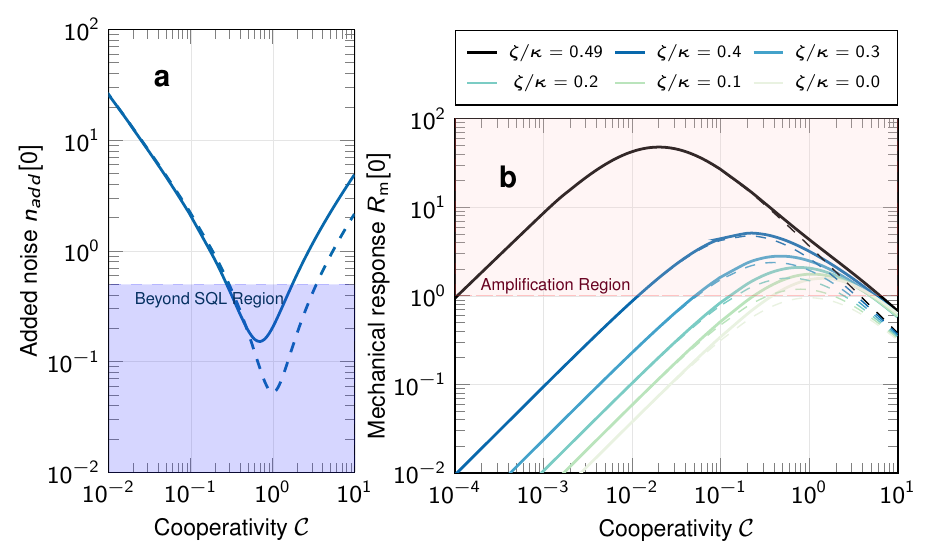}
		\caption{(a) The optically added noise of measurement and (b) the mechanical response as a function of the optomechanical cooperativity for different values of feedback gain and for fixed  $ \kappa_2/\kappa=0.95 $ and $ \kappa=2\omega_{\rm{m}} $. Solid lines and dashed lines correspond to beyond the RWA and RWA, respectively. Other parameters same as before. The noise suppression and the signal amplification regions are shown in blue and red, respectively.}	
		\label{Fig:Fig6}
	\end{figure}

	In Fig.~\ref{Fig:Fig6} (a), we show the optically added noise for a system with $ \kappa = 2\omega_{\rm m} $ to which RWA does not apply. The dashed lines correspond to the RWA in the system dynamics. One can see that the influence of the counter-rotating terms appears as we increase the optomechanical cooperativity. The minimum of the added noise increases by considering counter-rotating term and hence the system sensitivity decreases. In Fig.~\ref{Fig:Fig6} (b), we plot the mechanical response as a function of the optomechanical cooperativity for different values of feedback gain. We can also see that the counter-rotating terms contribute to the signal amplification in the system. Therefore, the price paid for signal amplification beyond RWA is sensitivity reduction.
\section{\label{Appendix:B} Scattering matrix elements}
The output and input fields are related via the scattering matrix $ {\bf s} $ as
\begin{align}
	{{\bf{n}}_{{\rm{out}}}^{(2)}}[\omega ] = {\bf{s}}[\omega ]{{\bf{n}}_{{\rm{in,fb}}}}[\omega ]-{{\bf{n}}_{{\rm{in}}}^{(2)}}[\omega ]\,,
\end{align}
where have defined $ {{\bf{n}}_{{\rm{in,fb}}}} = {[{c_{{\rm{in,fb}}}},c_{{\rm{in,fb}}}^\dag ,{b_{{\rm{in,f}}}},b_{{\rm{in,f}}}^\dag ]^T} $, $ {\bf{n}}_{{\rm{in}}}^{(2)} = {[c_{{\rm{in}}}^{(2)},c_{{\rm{in}}}^{(2)\dag },{b_{{\rm{in}}}},b_{{\rm{in}}}^\dag ]^T} $, $ {\bf{n}}_{{\rm{out}}}^{(2)} = {[c_{{\rm{out}}}^{(2)},c_{{\rm{out}}}^{(2)\dag },{b_{{\rm{out}}}},b_{{\rm{out}}}^\dag ]^T} $. Hereafter, the frequency dependence of these coefficients, represented by $ \omega $ is implicit for conciseness of notation. By introducing the susceptibilities of the cavity field and the mechanical oscillator as
	\begin{align}
		\chi _c^{ - 1}[\omega ] =&  - \kappa  + i\omega   + \zeta e^{ -i \phi } \, ,\\
		\chi _m^{ - 1}[\omega ] =& {-\gamma  + i\omega }\,,
	\end{align}
	with $ {\chi _c^*}[\omega ]=({\chi _c}[-\omega ])^* $ and $  {\chi _m^*}[\omega ]=({\chi _m}[-\omega ])^* $ and after straightforward calculations, we obtain the  matrix elements $ s_{ij}$ with $ i,j=1,\dots,4$  as  
	\begin{subequations}
		\begin{align}
			{s_{11}} =&  - 2\sqrt {{\kappa _2}{\kappa _{{\rm{fb}}}}} {\chi _c}\Xi \left( {{\Lambda ^2}\chi _c^*\chi _m^* + 1} \right)\, ,\\
			{s_{12}} =& 2\sqrt {{\kappa _2}{\kappa _{{\rm{fb}}}}} \zeta {e^{i\phi }}{\chi _c}\chi _c^*\Xi \, ,\\
			{s_{13}} =& 2i\sqrt {\gamma {\kappa _2}} {\chi _c}{\chi _m}\Lambda \Xi \left( {{\Lambda ^2}\chi _c^*\chi _m^* + 1} \right)\, ,\\
			{s_{14}} =& 2i\sqrt {\gamma {\kappa _2}} \zeta {e^{i\phi }}\Lambda {\chi _c}\chi _c^*\chi _m^*\Xi\, , \\
			{s_{21}} =& 2\sqrt {{\kappa _2}{\kappa _{{\rm{fb}}}}} \zeta {e^{ - i\phi }}{\chi _c}\chi _c^* \Xi \,  ,\\
			{s_{22}} =&  - 2\sqrt {{\kappa _2}{\kappa _{{\rm{fb}}}}} \chi _c^*\Xi \left( {{\Lambda ^2}{\chi _c}{\chi _m} + 1} \right)\, ,\\
			{s_{23}} =&  - 2i\sqrt {\gamma {\kappa _2}} \zeta {e^{ - i\phi }}\Lambda {\chi _c}\chi _c^*{\chi _m}\Xi \, ,\\
			{s_{24}} =&  - 2i\sqrt {\gamma {\kappa _2}} \chi _c^*\chi _m^*\Lambda \Xi \left( {{\Lambda ^2}{\chi _c}{\chi _m} + 1} \right)\, ,\\
			{s_{31}} =& 2i\sqrt {\gamma {\kappa _{{\rm{fb}}}}} {\chi _c}{\chi _m}\Lambda \Xi \left( {{\Lambda ^2}\chi _c^*\chi _m^* + 1} \right)\, ,\\
			{s_{32}} =&  - 2i\sqrt {\gamma {\kappa _{{\rm{fb}}}}} \zeta {e^{i\phi }}\Lambda {\chi _c}\chi _c^*{\chi _m}\Xi \, ,\\
			{s_{33}} =&   2\gamma {\chi _m}\Xi \left[ {\chi _c^*\left( {{\zeta ^2}{\chi _c} - {\Lambda ^2}\chi _m^*} \right) - 1} \right]\, ,\\
			{s_{34}} =& 2\gamma \zeta {e^{i\phi }}{\Lambda ^2}{\chi _c}\chi _c^*{\chi _m}\chi _m^*\Xi \, ,\\
			{s_{41}} =& 2i\sqrt {\gamma {\kappa _{{\rm{fb}}}}} \zeta {e^{ - i\phi }}\Lambda {\chi _c}\chi _c^*\chi _m^*\Xi \, ,\\
			{s_{42}} =&  - 2i\sqrt {\gamma {\kappa _{{\rm{fb}}}}} \chi _c^*\chi _m^*\Lambda \Xi \left( {{\Lambda ^2}{\chi _c}{\chi _m} + 1} \right)\, ,\\
			{s_{43}} =& 2\gamma \zeta {\Lambda ^2}{e^{ - i\phi }}{\chi _c}\chi _c^*{\chi _m}\chi _m^*\Xi \, ,\\
			{s_{44}} =&   2\gamma \chi _m^*\Xi \left[ {{\chi _c}\left( {{\zeta ^2}\chi _c^* - {\Lambda ^2}{\chi _m}} \right) - 1} \right]\, ,
		\end{align}
	\end{subequations}
	where we have defined 
	\begin{align}
		{\Xi ^{-1}} = {\Lambda ^2}\left( {{\Lambda ^2}{\chi _c}\chi _c^*{\chi _m}\chi _m^* + {\chi _c}{\chi _m} + \chi _c^*\chi _m^*} \right) + 1 - {\zeta ^2}{\chi _c}\chi _c^*\,,
	\end{align}
	Evidently, the output measurement quadrature of the field depends on the phase of the local oscillator, the feedback gain and the optomechanical coupling rate. In the case of zero optomechanical interaction $ \Lambda=0 $, the elements $s_{13}$, $s_{14}$, $s_{23}$, $s_{24}$, $s_{31}$, $s_{32}$, $s_{41}$ and $s_{42}$ become zero as one should expect.  
		
\section{\label{Appendix:C} Correlations in the frequency domain}
	Here, we transform correlations given by Eqs.~(\ref{Optical_correlation1})-(\ref{Optical_correlation4}) into the frequency-domain 
	\begin{align}
		\langle c_{\rm in,fb}[\omega ]c_{\rm in,fb}[\omega']\rangle  =& 2\pi \left( n_{\rm fb} - m_{\rm fb} \right)\delta (\omega  + \omega ')\,, \label{Optical_correlation1_frequency}\\
		\langle c_{\rm in,fb}^\dag [\omega ]c_{\rm in,fb}^\dag [\omega ']\rangle  =& 2\pi \left( n_{\rm fb} - m_{\rm fb}^* \right)\delta (\omega  + \omega ')\,,\label{Optical_correlation2_frequency}\\
		\langle c_{\rm in,fb}^\dag [\omega ] c_{\rm in,fb}[\omega ']\rangle  =& 2\pi n_{\rm fb}\delta (\omega  + \omega ')\,,\label{Optical_correlation3_frequency}\\
		\langle c_{\rm in,fb}[\omega ]c_{\rm in,fb}^\dag [\omega ']\rangle  =& 2\pi \big( n_{\rm fb} \!+\! \frac{\kappa }{{\kappa _{\rm fb}}} \!-\!   m_{\rm fb}\!-\! m_{\rm fb}^* \big)\delta (\omega  + \omega ')\,,\label{Optical_correlation4_frequency}
	\end{align}
\\
	We also use the following correlations in the derivation of the noise spectrum 
	\begin{align} 
			\langle {c_{{\rm{in}}}^{(2)}[\omega ]{c_{{\rm{in}},{\rm{fb}}}}[\omega ']} \rangle  &= 2\pi {p_{{\rm{fb}}}}\delta (\omega  + \omega ')\,,\\
			\langle {c_{{\rm{in}}}^{(2)}[\omega ]c_{{\rm{in,fb}}}^\dag [\omega ']} \rangle  &= 2\pi \left( {{p_{{\rm{fb}}}}+ \sqrt {{\kappa _{\rm{2}}}/{\kappa _{{\rm{fb}}}}} } \right)\delta (\omega  + \omega ')\,,\\
						\langle {c_{{\rm{in}}}^{(2)\dag } [\omega ]c_{{\rm{in,fb}}}^\dag [\omega ']} \rangle  &= \langle {c_{{\rm{in}}}^{(2)\dag } [\omega ]{c_{{\rm{in,fb}}}}(t')} \rangle  =0\,,\\
			\langle {{c_{{\rm{in,fb}}}}[\omega ]c_{{\rm{in}}}^{(2)\dag }[\omega ']} \rangle & = 2\pi \left( {p_{{\rm{fb}}}^*[\omega ] + \sqrt {{\kappa _{\rm{2}}}/{\kappa _{{\rm{fb}}}}} } \right)\delta (\omega  + \omega ')\,,\\
			\langle {c_{{\rm{in,fb}}}^\dag [\omega ]c_{{\rm{in}}}^{(2)\dag }[\omega ']} \rangle  &= 2\pi p_{{\rm{fb}}}^*\delta (\omega  + \omega ')\,,\\
			\langle {{c_{{\rm{in}},{\rm{fb}}}} [\omega ]c_{{\rm{in}}}^{(2)}[\omega ']} \rangle  &= \langle {c_{{\rm{in,fb}}}^\dag  c_{{\rm{in}}}^{(2)}[\omega ']} \rangle  = 0\,,
	\end{align}

\section{\label{Appendix:E} System stability}
	In this appendix, we use the Routh--Hurwitz criterion to investigate the stability of the system. The system reaches its steady state when all the eigenvalues of the drift matrix $ {\bf A}$ have negative real parts \cite{DeJesus}. The system is stable if the following conditions are satisfied: 
	\begin{align}
		{C_1} =& \gamma  - \zeta \cos \phi  + \kappa >0\,,\\
		{C_2} =& \zeta \cos \phi \left[ { - 4{\gamma ^2} - ({\cal C} + 8)\gamma \kappa  + 2\zeta (2\gamma  + \kappa )\cos \phi  - 3{\kappa ^2}} \right] \nonumber\\&+ (\gamma  + \kappa )\left[ {{\gamma ^2} + ({\cal C} + 3)\gamma \kappa  + {\kappa ^2}} \right] >0\,,\\
		{C_3} =&  \big\{ {\zeta \cos \phi \left[ {\zeta \cos \phi ({\cal C}\kappa  + 2\kappa  + 2\gamma ) - (\gamma  + \kappa )(2{\cal C}\kappa  + 3\kappa  + \gamma )} \right]} \big. \nonumber\\
		&\big. { + \kappa ({\cal C} + 1){{(\gamma  + \kappa )}^2}} \big\}(\gamma  - 2\zeta \cos \phi  + \kappa )>0 \,,\\
		{C_4} =& {\cal C}\kappa  - 2\zeta \cos \phi  + \kappa >0\,.
	\end{align}

\end{document}